%% file: main.tex
\documentclass[a4paper,pre,aps,twocolumn, reprint,  superscriptaddress, 10pt, longbibliography]{revtex4-2}
\usepackage[utf8]{inputenc}
\usepackage{amsmath,amsfonts,amssymb,bm}
\usepackage[mathscr]{euscript}
\usepackage{graphicx}
\usepackage[table,dvipsnames]{xcolor} 
\usepackage{tikz}
\usepackage{multirow} % Required for \multirow
\usepackage{soul}
\usepackage{makeidx}
\usepackage{changes}
\usepackage{todonotes}
\usepackage[colorlinks]{hyperref}
\hypersetup{
    colorlinks=true,
    linkcolor=NavyBlue,
    citecolor=ForestGreen,
    urlcolor=RoyalBlue,
    filecolor=Magenta,
    linktocpage=true
}
\def\bea{\begin{eqnarray}}
\def\eea{\end{eqnarray}}
\def\ba{\begin{array}}
\def\ea{\end{array}}
\def\n{\nonumber}

\def\la{\langle}
\def\ra{\rangle}

\begin{document}

\title{Non-monotonic diffusion from nonequilibrium driving}

\author{Manish Patel}
\email{manish.patel@iopb.res.in}
\affiliation{Institute of Physics, Sachivalaya Marg, Sainik School, Bhubaneswar 751005, India}

\affiliation{Homi Bhabha National Institutes (HBNI), Training School Complex, Anushakti Nagar, Mumbai, India 400094}

\author{Ritwick Sarkar}
\email{ritwick.sarkar@bose.res.in}
\affiliation{S. N. Bose National Centre for Basic Sciences, Kolkata 700106, India}
\author{Urna Basu}
\email{urna@bose.res.in}
\affiliation{S. N. Bose National Centre for Basic Sciences, Kolkata 700106, India}

\author{Debasish Chaudhuri}
\email[Cooresponding author:~]{debc@iopb.res.in}
\affiliation{Institute of Physics, Sachivalaya Marg, Sainik School, Bhubaneswar 751005, India}
\affiliation{Homi Bhabha National Institutes (HBNI), Training School Complex, Anushakti Nagar, Mumbai, India 400094}

\begin{abstract}

The stochastic dynamics of interacting particles far from equilibrium remains a fundamental challenge in statistical physics. While reciprocal interactions often permit effective one-body descriptions, such reductions generally fail for nonreciprocal interactions, which are ubiquitous in driven and active systems. We develop a unified theoretical framework for interacting particles with reciprocal and nonreciprocal couplings, applicable to the principal classes of active matter, including run-and-tumble, active Brownian, and active Ornstein–Uhlenbeck particles. As a minimal example, we study a passive particle driven by an active particle. On a periodic ring, we show analytically that the driven particle is always diffusive at long times, independent of the microscopic driving mechanism. Analytical predictions and numerical simulations reveal a nonmonotonic dependence of the effective diffusivity on the driving activity, giving rise to both enhanced and suppressed transport. Remarkably, the same behavior occurs in an equilibrium system driven out of equilibrium by coupling the driving particle to a higher local temperature. Our framework quantitatively captures both systems, identifies the common mechanism underlying the nonmonotonic transport, and establishes a unified description of transport under active and passive nonequilibrium driving.

\end{abstract}

\maketitle

\section{Introduction}

Understanding the stochastic dynamics of interacting particles remains a central challenge in statistical physics~\cite{liggett1985interacting,spohn2012large}. This challenge is amplified in nonequilibrium systems, where the absence of a Maxwell-Boltzmann description leads to rich and often unexpected behavior. A paradigmatic example is two interacting particles coupled to thermal reservoirs at different temperatures. Despite its simplicity, this system exhibits diverse phenomena, including synchronized spinning~\cite{SciPostPhysCore.6.3.056}, handed correlations~\cite{abdoli2026}, transient exchange fluctuation theorems~\cite{Prl_116_berut,PhysRevE.94.052148}, and classical analogues of quantum entanglement~\cite{PhysRevLett.134.227101}.

Another minimal nonequilibrium system is a pair of interacting active particles, where self-propulsion gives rise to dynamical behaviors absent in passive systems~\cite{RevModPhys.88.045006,annurev-conmatphys-031214-014710}. Activity can qualitatively modify interactions, inducing effective attraction independent of the underlying coupling~\cite{slowman2016jamming, twortp_lattice, rtp_dimer_hardcore,bound_state_rtp} and generating non-Gaussian steady-state distributions in harmonic traps~\cite{Chaudhuri_Dhar_2021, Patel_Chaudhuri_2024, Takatori2016}. Conversely, heterogeneity in self-propulsion speeds can produce emergent repulsion even when the microscopic interaction is attractive~\cite{D5SM00137D,Sarkar_2025}. An important question emerges: how does the dynamics change when a passive particle is coupled to an active one.

An effective two-particle description also emerges naturally from many-body systems in the presence of nonequilibrium environments~\cite{nphy_ybx6,ginot2025energy,PhysRevLett.91.248301,Hayashi_2006,PhysRevLett.128.028001}, which might even lead to the interaction between the particles being non-reciprocal~\cite{Mandal2024, Golestanian_2024, 101063_50289552,passive_driven_brownian}. Systems with non-reciprocal interactions have recently gained immense interest due to their unusual collective properties, both in active and passive systems~\cite{Paul_Chaudhuri_2026, nonreci_active_passsive,Saha_2019,PhysRevLett.123.018101,Loos_2020}. Even at the level of two-particle set-up, non-reciprocal interactions can give rise to a variety of intriguing phenomena, ranging from non-diffusive fluctuations at long-times~\cite{passive_driven_brownian}, non-trivial entropy production~\cite{Ganguly_Chaudhuri_2013, Chaudhuri_2014, Fodor2021, PhysRevE.100.050603}, to modification of escape probabilities \cite{PhysRevE.109.054129}. However, analytical study of non-reciprocal systems is harder even for a two-particle system due to the lack of an effective one-particle description.

A natural question is how nonequilibrium driving modifies the long-time transport of a passive particle. Here, we address this question using a one-dimensional model of a passive Brownian particle interacting with an active particle through a short-range potential. We focus on a run-and-tumble particle~\cite{FODOR2018106,solon2015active,Malakar_2018}, while the framework extends naturally to other active dynamics, including active Brownian particles~\cite{RevModPhys.88.045006,Shee2020,Chaudhuri_Dhar_2021} and active Ornstein--Uhlenbeck particles~\cite{PhysRevLett.117.038103,PhysRevE.103.032607}. We show that, independent of the reciprocity of the interaction, the effective diffusivity of the passive particle depends non-monotonically on the self-propulsion speed of the active particle. Remarkably, this behavior persists when the driving particle is passive but has a higher diffusivity. We further develop a unified analytical framework that captures this nontrivial transport behavior for both reciprocal and nonreciprocal interactions.

The paper is organized as follows. In the next section, we study an active–passive particle pair confined to a ring and interacting through a short-range potential. We derive the MSD of the passive particle in the weak-driving regime using a harmonic approximation and obtain the strong-driving behavior from the force autocorrelation generated by the active particle. In Sec.~\ref{passive_drive}, we extend the analysis to a passive driving particle with a different diffusivity. Finally, Secs.~\ref{discuss} and \ref{conclusion} summarize our findings, discuss their broader implications, and highlight future directions.

\begin{figure}[t]
    \centering
    \includegraphics[width=\linewidth]{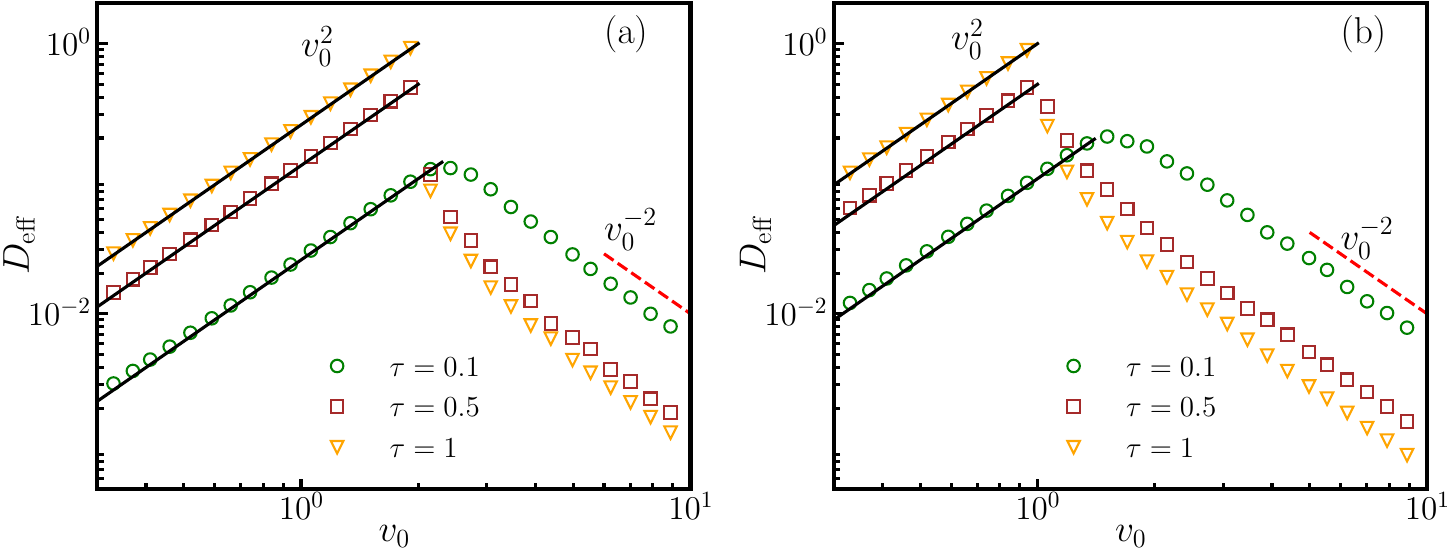}
    \caption{Behaviour of effective diffusion coefficient for $D=0$:  Plot of $D_\text{eff}$ as a function of self-propulsion speed $v_0$ for different persistence times $\tau$ for (a) the reciprocal case $\mu = 1$  and (b) for the non-reciprocal case $\mu = 0$. Here we have taken $L = 10$, $\lambda = 1$, and $k = 1$. The black solid lines correspond to the theoretical prediction Eq.~\eqref{eq:deff_harmonic_nrcp}. }  
     \label{fig:deff_full_range}
\end{figure}

\section{Active Driving}
We consider two interacting particles moving on a ring of circumference $L$. The particle with position $x(t)$ is passive and coupled to a thermal bath at temperature $T=\gamma D/k_B$, while the particle with position $y(t)$ is active and undergoes run-and-tumble (RTP) dynamics~\cite{FODOR2018106,solon2015active,Malakar_2018}. The equations of motion of the passive particle and the RTP are
\begin{align}
\dot{x}(t) &= -\frac{\partial}{\partial x}V(|x-y|)+\sqrt{2D}\,\eta(t), \qquad \quad\label{eq:eom1}\\
\dot{y}(t) &= -\mu\,\frac{\partial}{\partial y}V(|x-y|)+v_0\,\sigma(t),
\label{eq:ref_eom2}
\intertext{with}
V(z)&=k \lambda\,\exp{\left(-\frac{z}{\lambda}\right)}, \label{inter_pot}
\end{align}
where $\eta(t)$ is a Gaussian white noise with zero mean and unit variance, $v_0$ is the self-propulsion speed of the RTP, and $\sigma(t)$ represents a dichotomous process which switches between $\pm1$ with rate $\alpha$. The dichotomous noise $\sigma$ has a two-time correlation given by the following,
\bea
\la \sigma (t) \sigma(t')\ra = \exp{\left(-\frac{|t-t'|}{\tau}\right)},\quad {\rm where,}\quad \tau=\frac{1}{2 \alpha},\label{def_tau}
\eea 
and $\tau$ represents persistence time. 
This persistent time-dependent force correlation governs the MSD and is common to major models of self-propelled particles, including run-and-tumble particles, active Brownian particles, and active Ornstein–Uhlenbeck particles~\cite{Patel2025}. Therefore, the diffusivity results derived below are not restricted to RTPs but extend broadly across these classes of active matter models.

The parameter $\mu$ in Eq.~\eqref{eq:ref_eom2} controls the degree of reciprocity: $\mu=1$ corresponds to a conventional reciprocal interaction satisfying action–reaction symmetry, whereas $\mu=0$ represents the fully nonreciprocal limit, where the active particle drives the passive particle without experiencing any feedback.

%%%%%%%%%%%%%%%%%%%%%%%%%%
\begin{figure}[t]
    \centering
    \includegraphics[width=\linewidth]{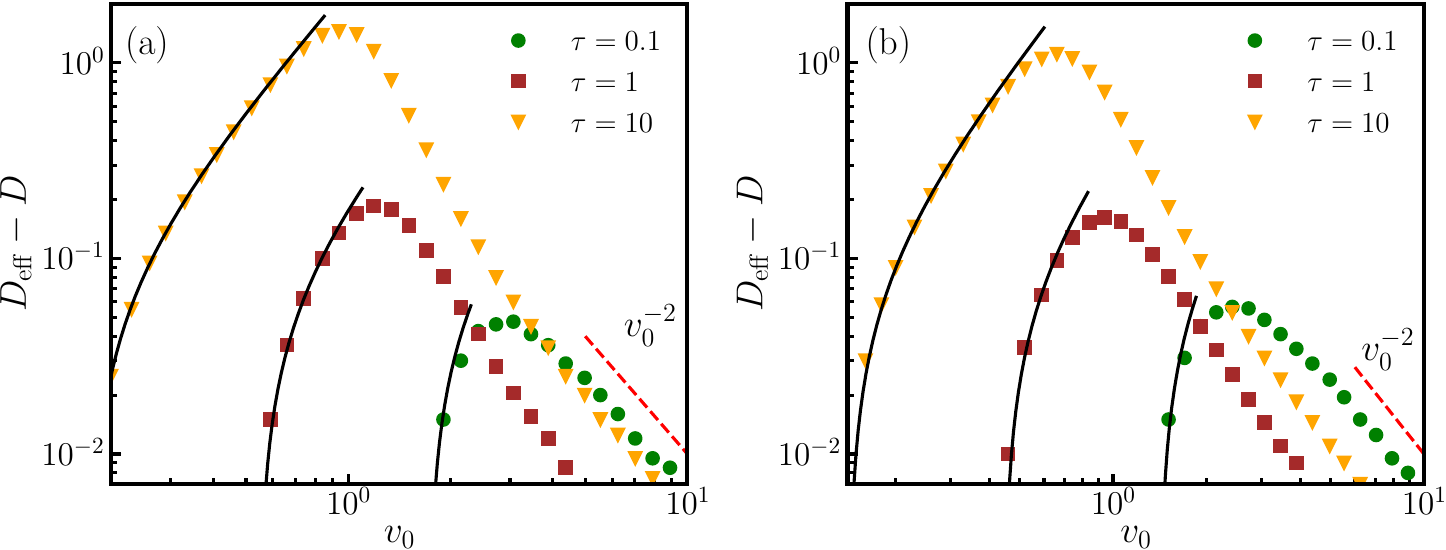}
    \caption{Behaviour of effective diffusion coefficient for $D>0$:  Plot of  $D_\text{eff}-D$ versus self-propulsion speed $v_0$ for different persistence times $\tau$ and  (a)  for reciprocal interaction $\mu=1$, and  (b) for non-reciprocal interaction with $\mu=0.5$, respectively. The black solid lines in both panels correspond to the analytical prediction Eq.~\eqref{effD_rtp} with respective values of $\mu$. The other parameters used are $D=0.1$, $L = 10$, $\lambda = 1$, and $k = 1$.}  
    \label{fig:deff_diffusion}
\end{figure}
%%%%%%%%%%%%%%%%%%%%%%%%%%%%%

To characterize the nonequilibrium transport induced by the active particle, we numerically integrate Eqs.~(\ref{eq:eom1}) and (\ref{eq:ref_eom2}) using an Euler–Maruyama discretization scheme. Throughout, we consider the short-range interaction potential $V(z)$ defined in Eq.~\eqref{inter_pot}. Although the dynamics are confined to a finite ring, the unfolded trajectory of the passive particle exhibits long-time diffusion for all self-propulsion speeds $v_0$, characterized by the effective diffusion coefficient
\begin{equation}
D_{\rm eff}
=\lim_{t\to\infty}
\frac{1}{2t}
\left[
\langle X_t^2\rangle-\langle X_t\rangle^2
\right],
\end{equation}
where $X_t$ denotes the unfolded position corresponding to the ring coordinate $x(t)$.

A central finding of our numerical study is that $D_{\rm eff}$ exhibits a nonmonotonic dependence on the self-propulsion speed $v_0$ of the driving particle, irrespective of the nonreciprocity parameter $\mu$ [see Figs.~\ref{fig:deff_full_range}–\ref{fig:deff_diffusion}]. Specifically, beyond an additive constant, the diffusivity increases quadratically with $v_0$ in the low-activity regime, reaches a maximum at intermediate $v_0$, and subsequently decreases as $v_0^{-2}$ at large $v_0$. The coexistence of transport enhancement and suppression driven by the same mechanism is striking. In the following, we develop an analytical theory that captures this nonmonotonic behavior and elucidates its underlying physical origin.

\subsection{Analytical theory of long-time diffusion}
\label{sec:scaling_v0}

In the low self-propulsion regime, where interactions dominate, the active particle remains within the interaction range and continuously pushes the passive particle, causing the pair to move together. Motivated by this bound-state picture, we approximate the interaction force harmonically and derive the passive-particle MSD in Sec.~\ref{sec_small_activity}.

In contrast, at large self-propulsion speeds, activity dominates the interaction, causing the active particle to repeatedly overtake the passive particle and deliver short force kicks during collisions. We describe these kicks as an effective stochastic force governed by the propagator of the interparticle separation. Using the resulting two-time force correlation, we derive the MSD of the passive particle in Sec.~\ref{sec_large_activity}.

\subsubsection{Small activity}\label{sec_small_activity}

In this regime, the persistent active force enhances the passive-particle diffusivity by an amount proportional to $v_0^2\tau$. This behavior is captured by a harmonic approximation of the interparticle interaction, yielding the Langevin dynamics
\bea
\dot{x}(t) &=&  k_{\rm eff} [a-(x-y)] + \sqrt{2 D} \, \eta(t),\label{eom_x_avg_pos} \\
\dot{y}(t) &=& - \mu \, k_{\rm eff} [a-(x-y)] + v_0 \sigma (t),\label{eom_y_avg_pos}
\eea
where, $a$ denotes the mean separation between the particles and $k_{\rm eff}$ characterizes the effective repulsion strength. For simplicity, Eqs.~(\ref{eom_x_avg_pos}) and~(\ref{eom_y_avg_pos}) assume that the particles remain in contact throughout the dynamics. Solving these equations [see Appendix~\ref{sec:small_v0_full_msd}] yields the steady-state MSD of the driven passive particle,
\bea
\label{eq:msd_harmonic_nrcp}
\la \Delta x^2 \ra (t) &=& 2 D_{\rm eff}t
- \frac{2 k_{\rm eff}^2}{k^{'3}_{\rm eff}} \left[ D + \frac{2 v_0^2 \alpha}{(4 \alpha^2 - k^{'2}_{\rm eff})} \right] (1 - e^{-k'_{\rm eff} \,t})\cr
&&  + \frac{v_0^2 k^2_{\rm eff}\left( 1 - e^{-2 \alpha t} \right)}{2 \alpha^2 (4 \alpha^2 - k^{'2}_{\rm eff}) } ,
\eea 
where $k^{'}_{\rm eff} = (1 + \mu) k_{\rm eff}$. At late time, MSD becomes diffusive with the effective diffusion constant $D_{\rm eff}$ as,
\bea
\label{eq:deff_harmonic_nrcp}
D_{\rm eff} &=& \frac{\mu^2 D}{(1+ \mu)^2} + \frac {v_0^2 \tau}{(1+ \mu)^2},\label{effD_rtp}
\eea
which depends only on the degree of non-reciprocity $\mu$ and is independent of the  interaction strength $k_{\rm eff}$. 

For reciprocal interactions ($\mu=1$), the effective diffusivity becomes $D_{\rm eff}=(D+v_0^2\tau)/4$, where $\tau$ is defined in Eq.~\eqref{def_tau}. In contrast, in the fully nonreciprocal limit ($\mu=0$), we obtain $D_{\rm eff}=v_0^2\tau$, showing that the diffusivity of the driven particle no longer depends on its bare diffusivity $D$.

Figures~\ref{fig:deff_full_range} and \ref{fig:deff_diffusion} show the comparison between simulations and theoretical predictions for both $D=0$ and $D>0$, and for reciprocal as well as nonreciprocal interactions. The close agreement across all cases demonstrates the accuracy of the harmonic approximation in the low-$v_0$ regime.

Interestingly, the same figures show two distinct trends: $D_{\rm eff}$ grows with $\tau$ at small $v_0$, but decreases with $\tau$ at large $v_0$. In the following section, we analytically investigate the large-$v_0$ behavior of $D_{\rm eff}$.

\subsubsection{Large activity}\label{sec_large_activity}

Beyond a critical self-propulsion speed, set by the interaction strength $k$, the active particle acquires sufficient energy to overcome the interaction barrier and repeatedly cross the passive particle. In this regime, collisions can be viewed as an effective stochastic force acting on the driven particle.
We model the interaction force as
\bea
f(|x-y|)=\frac{f_0}{\sqrt{2\pi\sigma_0^2}}e^{-(x-y)^2/2\sigma_0^2},
\label{interaction_force}
\eea
where $f_0$ and $\sigma_0$ characterize the force strength and range, respectively. These parameters depend on the microscopic interaction details, including reciprocity and system size, and thus effectively incorporate both reciprocal and nonreciprocal cases. This form, however, yields a nonzero long-time force correlation,
${\bar f}^{2}=\lim_{|t-t'|\to\infty}\langle f(t)f(t')\rangle$. Since collisions occur with equal probability from either side at long times, the mean force should vanish. We therefore consider the fluctuating collision force $\delta f=f(t)-\bar f$.

To obtain the propagator of the relative separation $r$, we neglect the interaction force as an approximation. The resulting Fokker–Planck equation yields the propagator [see Appendix~\ref{sec:propagator}].
\bea
 P(r,t|r_0,t') = \frac{1}{L} \sum_{n=-\infty }^{\infty} e^{i q_n (r - r_0) } \phi_n(|t-t'|),
 \eea
 with
\begin{align}
q_n = \frac{2 \pi n}{L},~
\omega_n = \sqrt{v_0^2 q_n^2 - \alpha^2},~a_n=\alpha+Dq_n^2,\label{q_n_omega_n_a_n}
\end{align}
and,
\begin{align}
\phi_n(s) = \exp{\left[-(\alpha +  D q_n^2)s\right]} \Big[ \cos{(\omega_n s)} +\frac{a_n}{\omega_n} \sin{(\omega_n s)}\Big].\label{phi_n}
\end{align}
In the long-time limit, the steady-state distribution $P_{\rm st}=1/L$. The two-time correlation can then be expressed in terms of the propagator as 
\begin{align}
\label{eq:force_corr_def}
\la f(t) f(t') \ra = \frac{f_0^2}{2 \pi \sigma_0^2} \int dr \int dr' \exp{\left(-\frac{r^2 + r'^2}{2\sigma_0^2}\right)} \cr \times P(r,t|r',t') P_{\rm {st}},
\end{align}
resulting in  $\la f(t) f(t') \ra = (f_0^2/L^2) \sum_{n = -\infty}^{\infty} \phi_n(|t-t'|) \, e^{-q_n^2 \sigma_0^2/2}$. Note that, in the long limit $t \to \infty$, only the $\phi_0$ term contributes to the propagator, which results in $\lim_{|t-t'| \to \infty} \la f(t) f(t') \rangle = f_0^2/L^2 \equiv {\bar{f}}^2$. Using these results, we obtain the steady-state two-time correlation of the effective force felt by the passive particle as,
\bea
\label{eq:force_corr_gauss}
\langle \delta f(t) \delta f(t') \rangle = \frac{2 f_0^2}{L^2} \sum_{\substack{n=1}}^{\infty} \phi_n(|t-t'|) \, e^{-q_n^2 \sigma_0^2/2}.
\eea

%%%%%%%%%%%%%%%%%%%%%%
\begin{figure}[t]
    \centering
    \includegraphics[width=\linewidth]{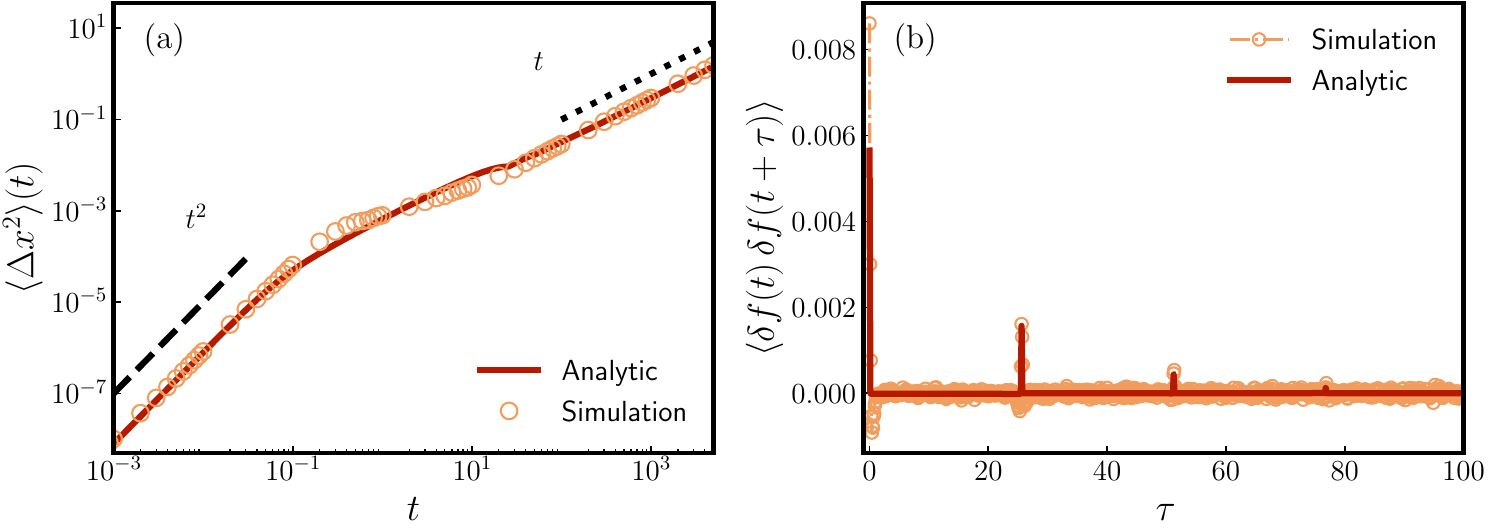}
    \caption{Mean-squared displacement and two-time force correlation as a function of time. The line plots in (a) and (b) are obtained from Eq.~(\ref{eq:msd_large_v0_spread}) and Eq.~(\ref{eq:force_corr_gauss}) with parameter fitting $f_0 = 0.66$, $\sigma_0 = 0.17$, and using $n=300$ modes, respectively. The points denote the result from the simulation obtained at parameter values $v_0 = 5$, $L = 128$, $\tau = 10$, $\mu = 1$, $k = 1$, $D=0$, and $\lambda = 1$.}
    \label{fig:msd_compare_large_v0}
\end{figure}
%%%%%%%%%%%%%%%%%%%%%%%%%%

The MSD of the passive driven particle can be written as, 
\bea
\langle \Delta x^2 \rangle(t)
= 2 D t + \int_{0}^t ds \int_{0}^{t} ds' \la \delta f(s) \delta f(s') \ra.
\eea
The above integral can be evaluated exactly using  Eq.~\eqref{eq:force_corr_gauss}, leading to the MSD,
\begin{align}
&\langle \Delta x^2 \rangle(t) =2Dt
+\frac{4f_0^2}{L^2}
\sum_{n=1}^{\infty}
e^{-\frac{\sigma_0^2 q_n^2}{2}}
\Bigg[\frac{2a_n\, t}{a_n^2+\omega_n^2}
+\frac{\omega_n^2-3a_n^2}
{(a_n^2+\omega_n^2)^2} \cr 
&+ e^{-a_n t} \left\{\frac{(3a_n^2-\omega_n^2)}{(a_n^2+\omega_n^2)^2}\cos\omega_n t +
\frac{a_n(a_n^2-3\omega_n^2)}{\omega_n(a_n^2+\omega_n^2)^2}
\sin\omega_n t \right\}
\Bigg], \label{eq:msd_large_v0_spread}
\end{align}
where $a_n$, $\omega_n$ and $q_n$ are as defined in Eq.~\eqref{q_n_omega_n_a_n}.
The analytical expression provides an excellent description of the numerical simulation results for the MSD, as demonstrated in Figs.~\ref{fig:msd_compare_large_v0} and \ref{fig_manyL}. In evaluating the expression, we include $n=300$ modes to achieve convergence of the series, while the parameters $f_0$ and $\sigma_0$ are extracted by fitting the short- and long-time regimes of the MSD, as presented below.

It is instructive to consider the behaviour of the MSD in the short- and long-time regimes separately.
In the short time limit, we expand the MSD around $t = 0$ to obtain
\begin{align}
\label{eq:largev_0_small_time}
\la \Delta x^2 \ra (t) \approx \frac{ 2f_0^2 t^2}{L^2} \sum_{n=1}^{\infty} e^{-\frac{\sigma_0^2 q_n^2}{2}} 
=\frac{f_0^2}{L^2} \left( \frac{L}{\sqrt{2 \pi}\, \sigma_0} - 1\right) t^2.
\end{align}
The MSD exhibits ballistic nature in this limit, with a drift that depends on the strength $f_0$ and range $\sigma_0$~[see Eq.~\eqref{interaction_force}].

In the long time limit, the MSD shows diffusive behavior $\la \Delta x^2\ra_{t \to \infty} = 2 D_{\rm eff} t$ with effective diffusivity 
\bea
\label{eq:full_deff_largev0}
D_{\rm {eff}} = D + \frac{2f_0^2}{L^2} \sum_{n=1}^{\infty} \frac{2 a_n}{a_n^2 + \omega_n^2} e^{-\sigma_0^2 q_n^2/2}. 
\eea
 Further, for the case of a stationary driven particle $D = 0$, the effective diffusion reduces to 
\bea
\label{eq:largev_0_deff}
D_{\rm {eff}} =\frac{2f_0^2}{L^2 v_0^2 \tau } \sum_{n=1}^{\infty} \frac{e^{-\sigma_0^2 q_n^2/2}}{q_n^2}  
\approx \frac{f_0^2}{12  v_0^2 \tau}, 
\eea
in the limit of small interaction range $\sigma_0 /L \ll 1$. In general, $D_{\rm {eff}}-D \sim v_0^{-2}$, as shown in Fig.~\ref{fig:deff_full_range} and \ref{fig:deff_diffusion} for large self-propulsion speed $v_0$.

Using the short- and long-time MSD behaviors from Eqs.~(\ref{eq:largev_0_small_time}) and (\ref{eq:full_deff_largev0}), we estimate the effective force scale $f_0$ and range $\sigma_0$. With these parameters, the analytical MSD in Eq.~(\ref{eq:msd_large_v0_spread}) agrees well with simulations in both the short- and long-time limits while capturing the crossover between them [Figs.~\ref{fig:msd_compare_large_v0} and \ref{fig_manyL}]. Using the same values of $f_0$ and $\sigma_0$, Eq.~(\ref{eq:force_corr_gauss}) also reproduces the two-time force correlation in good agreement with simulations [Figs.~\ref{fig:msd_compare_large_v0}(b) and \ref{fig_manyL}(b)–(d)].

\subsection{System size dependence and single-file diffusion}
\subsubsection{Effect of system size}

\begin{figure}[t]
    \centering
    \includegraphics[width=\linewidth]{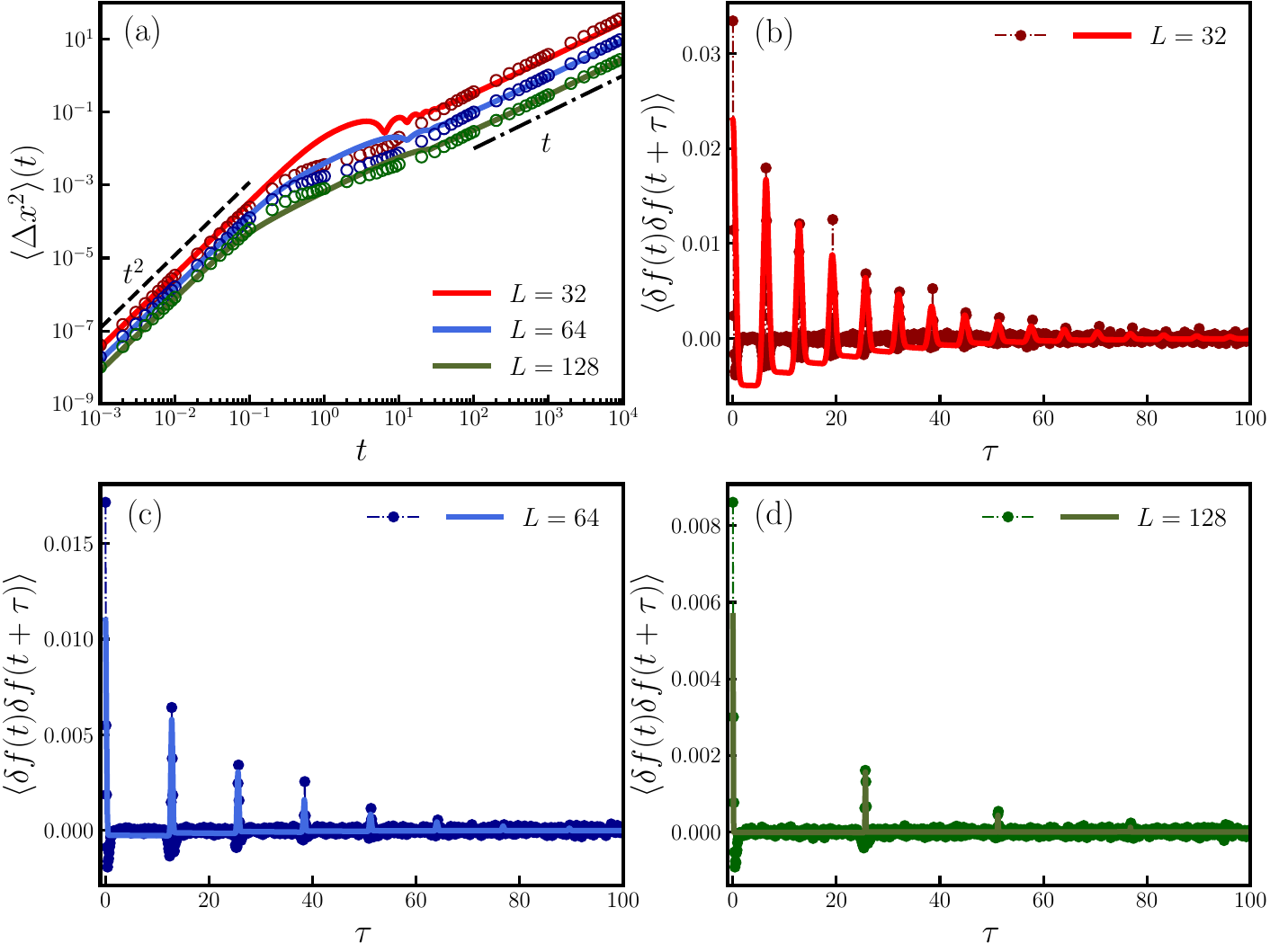}
    \caption{Mean-squared displacement and two-time force correlation. The solid lines in (a) are obtained from Eq.~\eqref{eq:msd_large_v0_spread}, while those in (b)–(d) follow Eq.~\eqref{eq:force_corr_gauss} using fitted parameters $f_0=2.44$, $\sigma_0=1.82$ for $L=32$, $f_0=1.22$, $\sigma_0=0.58$ for $L=64$, and $f_0=0.66$, $\sigma_0=0.17$ for $L=128$. Both expressions are evaluated using $n=300$ modes. Symbols denote simulations with $v_0=5$, $\tau=10$, $\mu=1$, $k=1$, $D=0$, and $\lambda=1$.}
    \label{fig_manyL}
\end{figure}

In the large-$v_0$ regime, the effective kick picture becomes increasingly accurate as the system size exceeds the interaction range and collisions become infrequent. Consistent with this, the analytical MSD in Eq.~\eqref{eq:msd_large_v0_spread} shows progressively better agreement with simulations as $L$ increases from $32$ to $128$ [Fig.~\ref{fig_manyL}(a)]. The longer time between collisions also reduces the periodicity and amplitude of the effective force correlation, as predicted by Eq.~\eqref{eq:force_corr_gauss} and confirmed by simulations [Fig.~\ref{fig_manyL}(b)–(d)].

%%%%%%%%%%%%%%%%%%%%%%%%%%%%%%%%%%%%%%%%%
%%%%%%%%%%%%%%%%%%%%%%%%%%%%%%%%%%%%%%%%%
\subsubsection{Single-file diffusion as intermediate scaling in large systems}
\begin{figure}[t]
    \centering
    \includegraphics[width=1\linewidth]{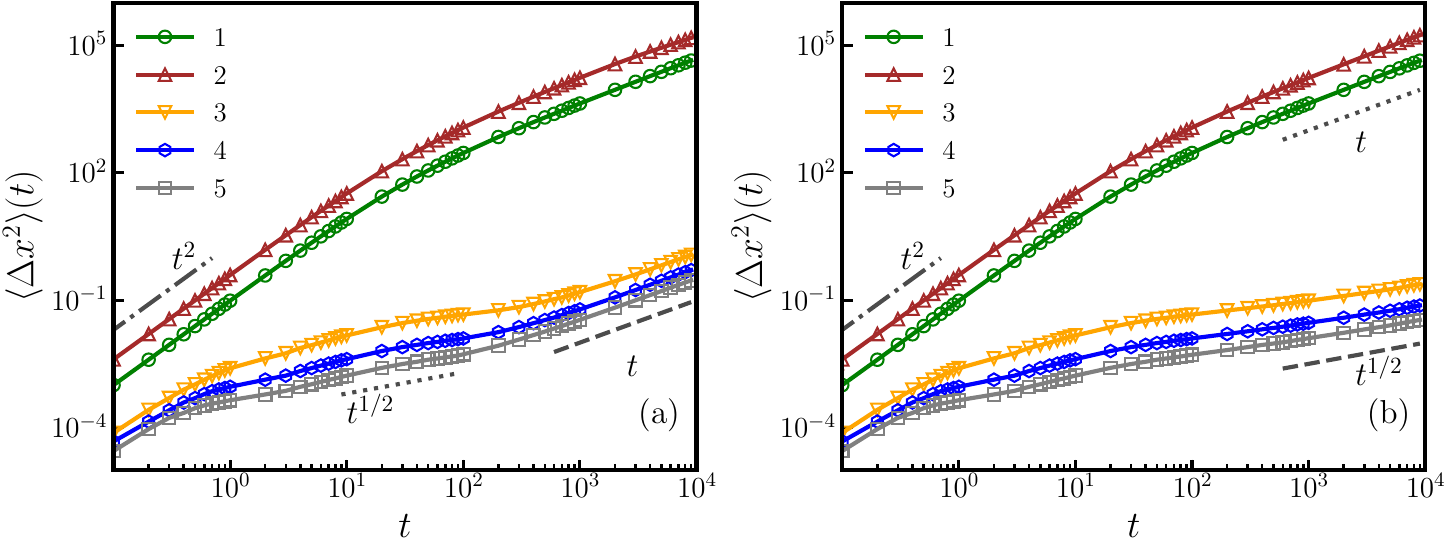}
    \caption{Active driving: Mean-squared displacement $\la \Delta x^2 \ra(t)$ versus time $t$ for (a) periodic boundary in unwrapped coordinates, and (b) open boundary. The parameter values are $\lambda = 1, k = 1, L = 512, \tau = 10, D = 0$. The inset labels represent the value of self-propulsion speed $v_0$, and the plots are obtained from numerical simulation.}
    \label{fig:msd_pbc_vs_free}
\end{figure}

At short times, persistent pushing by the active particle drives ballistic motion of the passive particle, $\langle \Delta x^2 \rangle(t)\sim t^2$, independent of the boundary conditions. At large activity, however, the late-time dynamics become boundary dependent. Under open boundaries, the active particle delivers intermittent kicks separated by first-passage times on an unbounded domain, leading to single-file diffusion (SFD) of the passive particle, $\langle \Delta x^2\rangle(t)\sim \sqrt{t}$ [Fig.~\ref{fig:msd_pbc_vs_free}(b)], consistent with Ref.~\cite{passive_driven_brownian}. Under periodic boundaries, the same SFD regime emerges only at intermediate times for sufficiently large systems, where the active particle has not yet explored the entire domain [Fig.~\ref{fig:msd_pbc_vs_free}(a)]. At longer times, finite-size effects restore normal diffusion, consistent with the small- and large-$v_0$ limits discussed in Sec.~\ref{sec:scaling_v0}.

%%%%%%%%%%%%%%%%%%%%%%%%%%%%
%%%%%%%%%%%%%%%%%%%%%%%%%%%%%%%%%%%

\section{Passive Driving}\label{passive_drive}

In this section, we consider a passive Brownian particle with diffusivity $D=k_B T/\gamma$ driven by another Brownian particle with higher diffusivity $D_0=k_B T_0/\gamma$ ($D_0>D$). Denoting their positions by $x$ and $y$, respectively, the dynamics are governed by the coupled Langevin equations:
\bea
\dot{x}(t) &=& - \frac{\partial}{\partial x} V(|x-y|) + \sqrt{2D} \, \eta^x(t), \\ 
\dot{y}(t) &=&- \mu \frac{\partial}{\partial y} V(|x-y|) + \sqrt{2D_0}\, \eta^y (t), 
\eea
where $\eta^x (t)$ and $\eta^y(t)$ are independent Gaussian white noises with zero mean and auto-correlations $\la \eta^x(t) \eta^{x}(t') \ra = \la \eta^y(t) \eta^{y}(t') \ra = \delta (t- t')$.

\subsection{Low diffusivity driving}

\begin{figure}[t!]
    \centering
    \includegraphics[width=1\linewidth]{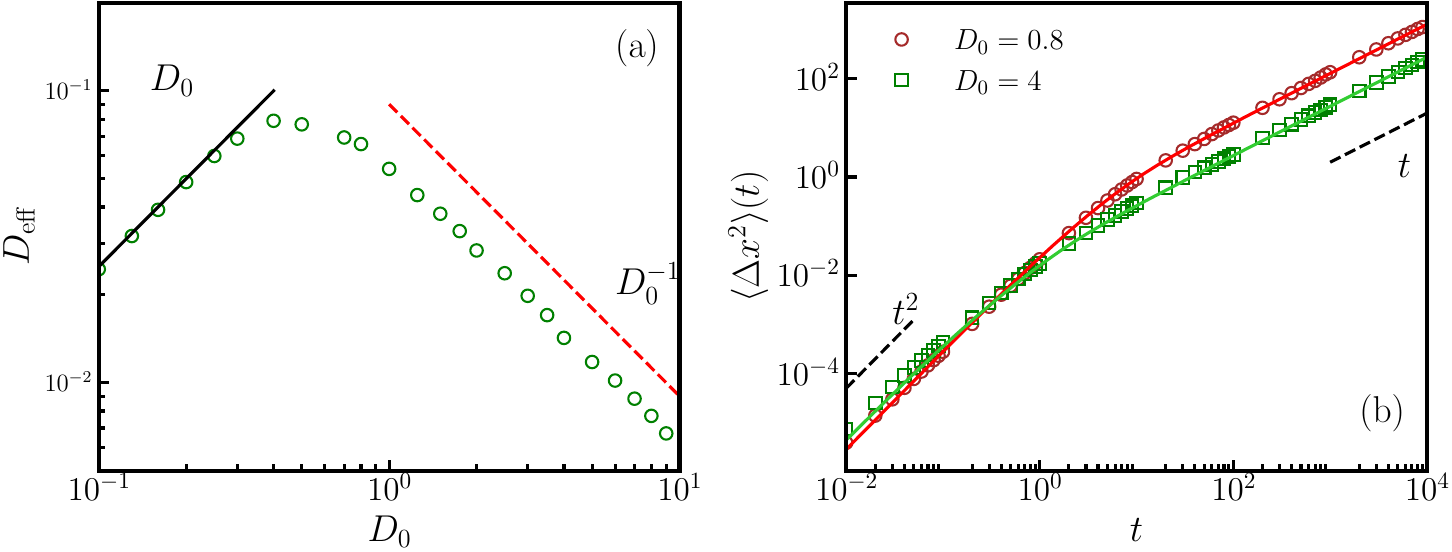}
    \caption{Passive driving. (a) Effective diffusivity of the driven particle versus the diffusivity of the driving particle for reciprocal interactions ($\mu=1$) with $D=0$. The solid black line denotes Eq.~\eqref{eff_D_thermal}, while the red dashed line shows the $D_0^{-1}$ scaling. (b) MSD of the driven particle for two values of $D_0$ with $D=0$. Solid lines are obtained from Eq.~\eqref{eq:msd_largeD0} using $n=300$ modes and fitted parameters $f_0=1.086$, $\sigma_0=1.162$ for $D_0=0.8$, and $f_0=0.971$, $\sigma_0=0.681$ for $D_0=4$. Parameters: $L=10$, $\lambda=1$, and $k=1$.}
    \label{fig:passive_driving}
\end{figure}

In the weak-driving limit, where the diffusivity contrast is small ($D_0 \gtrsim D$), the particles remain effectively noncrossing and move together during collisions, analogous to the active driving case. We therefore adopt the same harmonic approximation, assuming they remain in contact throughout the interaction. The Langevin equations then become:
\bea
\dot{x}(t) &=&  k_{\rm eff} \left[ a- (x(t)-y(t))\right] + \sqrt{2 D} \, \eta^x(t), \\
\dot{y}(t) &=& - \mu k_{\rm eff} \left[a- (x(t)-y(t))\right] + \sqrt{2 D_0} \eta^y (t),
\eea
where $a$ represents the mean separation between the two particles and $k_{\rm eff}$ is the strength of repulsion. Following the same procedure as for the active system in Appendix~\ref{sec:small_v0_full_msd}, we obtain the MSD of the driven passive particle as:
\bea
\la \Delta x^2 \ra (t) &=& 2 D_{\rm eff} t
+ \frac{2 k^2_{\rm eff}}{k^{'3}_{\rm eff}} (D_0 + D) (e^{-k'_{\rm eff}t} - 1),~~ 
\eea
with $k'_{\rm eff} = (1 + \mu) k_{\rm eff}$. The late-time MSD is diffusive, with effective diffusion coefficient
\bea
D_{\rm {eff}}&=& \frac{\mu^2 D}{(1+\mu)^2} + \frac{D_0}{(1+\mu)^2}. 
\label{eff_D_thermal}
\eea
The effective diffusion depends only on the degree of nonreciprocity $\mu$ and is independent of the interaction strength $k$, similar to the active driving case. For reciprocal interactions ($\mu=1$), we recover $D_{\rm eff}=(D_0+D)/4$, which agrees well with simulations [Fig.~\ref{fig:passive_driving}]. In the fully nonreciprocal limit ($\mu=0$), the effective diffusion becomes $D_{\rm eff}=D_0$, independent of the bare diffusivity $D$.

%%%%%%%%%%%%%%%%%%%%%%%%%%%%%%%%%%%%%%%%%%%%%%%%
%%%%%%%%%%%%%%%%%%%%%%%%%%%%%%%%%%%%%%%%%%%%%%%%

\subsection{High diffusivity driving}

In the limit of large driving-particle diffusivity, thermal fluctuations are sufficient to overcome the interaction barrier. We adopt the same physical picture as in the active case and consider the fluctuating force experienced by the passive particle, $\delta f=f(t)-\bar{f}$, with $f(t)$ given by Eq.~\eqref{interaction_force}. To obtain the propagator of the relative separation $r$, we neglect the interaction force as an approximation, reducing the dynamics to $\dot{r}=\sqrt{2D}\,\eta^x-\sqrt{2D_0}\,\eta^y$. Following the same procedure as for the active case, the corresponding Fokker–Planck equation becomes $\partial_t P=(D+D_0)\partial_r^2P$, yielding the propagator: 
\begin{align}
P(r,t|r_0,t') = \frac1L \sum_{n = -\infty}^{\infty} e^{-(D+D_0) q_n^2 |t-t'|} e^{i q_n (r-r_0)},
\end{align} 
with $q_n = 2 \pi n/L$.
Using Eq.~(\ref{eq:force_corr_def}), we obtain the two-time correlation of the effective force acting on the driven particle as,
\begin{align}
\la \delta f(t) \delta f(t') \ra = \frac{2f_0^2}{L^2} \sum_{n=1}^{\infty} e^{-(D+D_0) q_n^2 |t-t'|} e^{-q_n^2 \sigma_0^2/2}, \label{eq:force_thermal}
\end{align}
and the MSD of the driven particle as,
\begin{align}
\la \Delta x^2 \ra (t) &= 2 D t + \frac{4 f_0^2}{L^2} \sum_{n=1}^{\infty}
e^{ -\frac{\sigma_0^2 q_n^2}{2}} \Bigg[\frac{t}{q_n^2 (D_0+D)}  \cr 
& \qquad  + \frac {e^{-(D_0 + D)q_n^2 t}-1}{q_n^4 (D_0+D)^2 }  \Bigg]. \label{eq:msd_largeD0}
\end{align}

We compare the above expression with numerical simulations using the fitted values of $f_0$ and $\sigma_0$, finding good agreement [see Fig.~\ref{fig:passive_driving}]. At short times, the MSD exhibits ballistic scaling, $\langle \Delta x^2\rangle \approx \frac{f_0^2}{L^2}\left(\frac{L}{\sqrt{2\pi}\,\sigma_0}-1\right)t^2$. At late times, the MSD becomes diffusive, with effective diffusivity of the driven particle given by 
\begin{align}
D_{\rm {eff}} & = D + \frac{2f_0^2}{L^2(D_0 + D)} \sum_{n=1}^{\infty} \frac{e^{-\sigma_0^2 q_n^2/2}}{ q_n^2} \cr 
& \approx  D + \frac{f_0^2}{12(D_0 + D)},
\end{align}
in the limit $\sigma_0/L \ll 1$. This reveals a slowdown of the late-time effective diffusivity of the driven particle as $D_0^{-1}$. Moreover, the passive results can be directly recovered from the active case discussed in the main text by taking the limit $v_0^2/(2\alpha)=D_0$ with $(v_0,\alpha)\rightarrow\infty$.

\section{Discussion}\label{discuss}

The non-monotonic dependence of the driven-particle diffusivity on the nonequilibrium drive originates from a crossover between two distinct dynamical regimes. At weak driving, the interaction confines the particles into a bound pair that moves approximately as a single object. The driven particle inherits the fluctuations of the driving particle, leading to enhanced diffusion with increasing activity or temperature. In this regime, the pair behaves as a rigid rod with center-of-mass coordinate $r_{\rm com}=(\mu x+y)/(1+\mu)$, where nonreciprocity controls how strongly the driving fluctuations are transmitted.

At strong driving, the bound-state picture breaks down as the driving particle repeatedly crosses the driven particle. The dynamics are then governed by intermittent collisions separated by long independent excursions. The relevant timescale is the first-passage time of the driving particle around the ring, $t_L\sim L^2/D_a$, with $D_a=v_0^2\tau$ for active driving. Modeling collisions as correlated kicks with correlation time $t_L$ gives $D_{\rm eff}\propto t_L\propto v_0^{-2}$. Thus, although individual collisions become stronger with increasing activity, their decreasing frequency and correlation time suppress the long-time transport.

This first-passage mechanism also applies to thermal driving, where $t_L\sim L^2/D_0$ and consequently $D_{\rm eff}\sim D_0^{-1}$. Hence, active and thermal driving share the same underlying transport mechanism, despite their different microscopic origins.

\section{Conclusion}\label{conclusion}

We studied the dynamics of a passive Brownian particle driven by either an active particle or a high-temperature passive Brownian particle on a one-dimensional ring, allowing both reciprocal and nonreciprocal interactions within a unified framework. In all cases, the driven particle exhibits late-time diffusion with an effective diffusivity that depends non-monotonically on the strength of the nonequilibrium drive.

Combining analytical approaches for weak and strong driving regimes, we show that the initial enhancement and subsequent suppression of diffusion arise from a crossover between bound-pair motion and intermittent collision-driven dynamics. The weak-driving regime is described by a harmonic approximation of the interaction, while the strong-driving regime is governed by force correlations set by the first-passage dynamics of the driving particle. These approaches quantitatively capture the simulation results and predict the asymptotic scalings $D_{\rm eff}\sim(v_0^2\tau)^{-1}$ for active driving and $D_{\rm eff}\sim D_0^{-1}$ for thermal driving.

Our results highlight a general mechanism by which nonequilibrium driving can both enhance and suppress transport depending on the interplay between collision strength and temporal correlations. Extending this framework to many-body systems, where collective effects and crowding may generate qualitatively different transport phenomena, remains an important direction for future studies.

\section{ACKNOWLEDGMENTS}
M.P. acknowledges SAMKHYA high-performance computing cluster and other computational facilities at the Institute of Physics, Bhubaneswar. R.S. acknowledges support from the Council of Scientific and Industrial Research, India [Grant No. 09/0575(11358)/2021-EMR-I]. D.C. acknowledges financial support from the Department of Atomic Energy (DAE) through Grant No. 1603/2/2020/IoP/R\&D-II/15028, a Visiting Professorship at CY Cergy Paris Universit{\'e}, and an Associateship of IIT Bombay.

\appendix

\section{Small activity}
\label{sec:small_v0_full_msd}
In the regime of small activity, we approximate the interaction between a passive and an active particle as a harmonic repulsion having the dynamics given in Eqs.~\eqref{eom_x_avg_pos}-\eqref{eom_y_avg_pos}. We define $u = x -y -a$ to rewrite the dynamics as
\bea
\dot{x}(t) &=& - k_{\rm eff} u(t) + \sqrt{2 D} \eta (t),\\
\dot{u}(t) &=& - k'_{\rm eff} u (t) + \sqrt{2 D} \eta (t)- v_0 \sigma(t),\label{eq:u_dynamic}
\eea
where $k'_{\rm eff} = (1 + \mu) k_{\rm eff}$
Thus, the steady state MSD of the driven passive particle can be written as
\begin{align}
\label{eq:tracer_msd}
\la \Delta x^2 \ra(t)  = \int_{0}^{t} dt_1 \int_0^t dt_2 \Big[ k^2_{\rm eff} \la u(t_1) u(t_2) \ra + 2 D \la \eta(t_1) \eta(t_2) \ra  \cr
  - k_{\rm eff} \sqrt{2D} \la u(t_1) \eta(t_2) \ra - k_{\rm eff} \sqrt{2D} \la \eta(t_1) u(t_2 )\ra \Big].
\end{align}
The steady state correlations can be obtained by solving Eq.~(\ref{eq:u_dynamic}), which take the form
\bea
\la u(t_1) u(t_2) \ra &=& \left( \frac{D}{k'_{\rm eff}} + \frac{2 v_0^2 \alpha}{ k'_{\rm eff} ( 4 \alpha^2  - k^{'2}_{\rm eff})} \right) e^{-k'_{\rm eff} |t_1-t_2|} \n\\
&-& \frac{v_0^2}{ (4 \alpha^2 - k^{'2}_{\rm eff})} e^{-2 \alpha |t_1-t_2|}, \\
\la u(t_i) \eta(t_j) \ra &=& \frac{\sqrt{2D}}{k'_{\rm eff} }\delta(t_i - t_j).
\eea
Using the above correlations along with Eq.~(\ref{eq:tracer_msd}), one arrives at the steady state value of the MSD given by Eq.~\eqref{eq:msd_harmonic_nrcp} in the main text. \\
 
\section{Propagator for relative coordinate}
\label{sec:propagator}
To simplify the propagator calculation, we neglect the collision term for relative-coordinate dynamics, leading to $\dot r = v_0 \sigma - \sqrt{2D}\, \eta$ where $r= (x-y)$. The corresponding Fokker-Planck equation follows 
\bea
\partial_t P_+ &=& - v_0 \partial_r P_+ + D \partial_r^2 P_+ - \alpha P_+ + \alpha P_-, \\
\partial_t P_- &=& v_0 \partial_r P_- + D \partial_r^2 P_- - \alpha P_- + \alpha P_+.
\eea
To solve the Fokker-Planck equation, we define the densities $P = P_+ + P_-$ and $Q = P_+- P_-$ and write
\begin{align}
\label{eq:prob_dynamics}
\partial_t^2 P + 2 \alpha \partial_t P = v_0^2 \partial_r^2P + 2 \alpha D \partial_r^2P + 2 D \partial_r^2 \partial_t P-D^2 \partial_r^4 P,
\end{align}
which are subjected to initial conditions $P(r,0|r_0,0) = \delta(r - r_0)$ and $\partial_t P(r,0|r_0,0) = 0$. 
Using periodic boundary conditions, we expand the propagator in Fourier modes as 
\begin{align}
P(r,t|r_0,0) = \frac{1}{L} \sum_{n = -\infty}^{\infty} \phi_n(t) e^{i q_n (r - r_0)},
\end{align}
with $q_n  = 2 \pi n/L$, where $n \in \mathbb{Z}$. 
We use the above expression in Eq.~(\ref{eq:prob_dynamics}) with initial conditions to obtain $\phi_n(t)$. The explicit expression of $\phi_n(t)$ is given by,
\bea
\phi_n (t) = e^{- (\alpha + D q_n^2 )t} \left[ \cos(\omega_n t) + \frac{\alpha + D q_n^2}{\omega_n} \sin(\omega_n t) \right],~~
\eea
where $\omega_n = \sqrt{v_0^2 q_n^2 - \alpha^2}$.

\bibliographystyle{apsrev4-2}
%\bibliography{ref}
\input{main.bbl}
\end{document}